\documentclass[aps,pra,twocolumn,groupedaddress,10pt]{revtex4-1}

\usepackage{amsmath}
\usepackage{graphicx}
 \usepackage{color}

\begin{document}
\title{Bound states with complex frequencies near the continuum on
  lossy periodic structures} 

\author{Zhen Hu}
\affiliation{Department of Mathematics, Hohai University, Nanjing,
  Jiangsu, China}
\author{Lijun Yuan}
\affiliation{College of Mathematics and Statistics, Chongqing Technology and Business University, Chongqing, 
China}
\author{Ya Yan Lu}
\email{Corresponding author: mayylu@cityu.edu.hk}
\affiliation{Department of Mathematics, City University of Hong Kong,
  Hong Kong, China}

\begin{abstract}
On a lossless periodic dielectric structure sandwiched between two
homogeneous media,  bound states in the continuum (BICs)
with real frequencies  and real Bloch wavevectors may exist, and they
decay exponentially in the surrounding homogeneous media and do not
couple with propagating plane waves with the same frequencies
and wavevectors. The BICs are of significant
current interest, because they give rise to high-$Q$ resonances when
the structure or the Bloch wavevector is slightly perturbed. In this
paper, the effect of a small material loss on the BICs 
  is analyzed by a perturbation method and illustrated by numerical
  results. 
It is shown
that bound states with complex 
frequencies near the continuum appear, but they behave
differently depending on whether the BIC is symmetry-protected or
not. The Bloch wavevector of  a bound state with a
complex frequency can be real if the original BIC is
symmetry-protected, and it is usually complex if the original BIC is not 
symmetry-protected.  Our study improves the theoretical understanding on
BICs and provides useful insight for their practical applications. 
\end{abstract}

\maketitle

\section{Introduction}

There is currently a significant 
research interest on bound states in the continuum (BICs) 
 in the photonics community \cite{hsu16,kosh19}. 
A BIC is a trapped or guided mode with
a frequency in the radiation continuum \cite{neumann29}. Although a BIC has the same
frequency (and wavevector when appropriate) as the radiative waves, it
does not couple with these waves and does not leak power to
infinity. Some early examples of BICs are trapped modes in waveguides
with local distortions
\cite{urse87,evans91,evans94}, where the
radiative waves are propagating 
modes of the waveguide without the local distortion. BICs are also
studied on uniform waveguides with lateral leaky structures
\cite{longhi07,plot11,weim13,zou15}. In that
case, a BIC is guided mode of the waveguide and the radiative waves
are propagating waves on the lateral structure. A particularly simple
structure supporting BICs is a slab waveguide with anisotropic core
and substrate \cite{shipman13,gomis17}. 
Many recent works are concerned with BICs on periodic structures,
including two-dimensional (2D) structures with one periodic direction 
\cite{bonnet94,shipman03,shipman07,hu15,port05,mari08,hsu13_1,bulg14b,gao16,yuan17,hu17josab,yuan17_4,bulg17pra,hu18,doel18},
three-dimensional (3D) biperiodic structures
\cite{padd00,tikh02,lee12,hsu13_2,yang14,zhen14,gan16,li16,ni16,zhang18}, and
3D rotationally symmetric structures
\cite{bulg15,bulg17}.  In these studies, the periodic structures are 
sandwiched between or surrounded by homogeneous media, the BICs are
guided Bloch modes above the lightline,  and the radiative waves are
propagating plane waves in the homogeneous media.

Importantly, a BIC can be regarded as a resonant mode (or resonant
state) with an infinite quality factor ($Q$-factor). This implies that
resonant modes with arbitrarily
high $Q$-factors  
can be created  by modifying the structure \cite{kosh18}  or varying a
physical parameter such as a component of the Bloch wavevector \cite{lijun17pra,lijun18pra}. Applications
of optical BICs include lasing \cite{kodi17}, sensing \cite{yesi19}, filtering
\cite{foley14,cui16}, switching \cite{han19}, nonlinear optics \cite{lijun16pra,lijun17pra}, 
etc. They are mostly related to the high-$Q$ resonances, and are based on 
local fields enhanced by the resonances \cite{mocella15,yoon15}, or special features 
in scattering, transmission or other spectra
\cite{shipman05,shipman12}. The existence  
of a BIC implies that the associated scattering or diffraction problem loses
uniqueness \cite{bonnet94,shipman07}. The non-uniqueness may be explored for potential
applications in optical switching and nonlinear optics
\cite{lijun16pra}. 

For theoretical understanding and potential applications, it is crucial to study
how a BIC is affected by a small perturbation of the structure or a
physical parameter. A BIC can be robust (i.e. it continues to exist) to a
special family of perturbations,  but it usually turns to a resonant
mode under a generic perturbation. Some BICs are symmetry-protected in the
sense that they exhibit a symmetry mismatch with the corresponding
radiative waves \cite{evans94,plot11,bonnet94,shipman03,lee12}. These BICs are usually found on structures with
certain symmetry, and  they are robust to
perturbations preserving the required symmetry \cite{bonnet94,shipman07}.
There are BICs that are
not protected by symmetry in the usual sense (i.e. there is no
symmetry mismatch), but depend crucially on symmetry for their
existence and robustness  \cite{zhen14,bulg17pra,yuan17_4}. 
If a BIC is turned to a resonant mode under a perturbation, it is 
important to find a relation between the $Q$-factor
and the magnitude of the perturbation \cite{kosh18,lijun17pra,lijun18pra}. 

Strictly speaking, an optical BIC only exists on a lossless dielectric
structure given by a real dielectric function (or tensor) $\epsilon$. Therefore, to
study the robustness of a BIC, one assumes  that the
perturbation to $\epsilon$ is also real \cite{yuan17_4}. Since actual materials are always
lossy, it is important to analyze the effect of a small material 
loss to the BIC. For a symmetry-protected BIC, if the absorption profile keeps the
symmetry of the structure, the BIC should remain as a bound
state uncoupled with the radiative waves, but the frequency of
the bound state becomes complex. In general, a BIC  unprotected by
symmetry 
is likely to become a resonant mode that radiates power to
infinity, and also loses power to dissipation. 
But are there special class of BICs (unprotected by
symmetry) that remain as bound states (without radiating power to
infinity) when a small material loss is added to the structure? In this paper, we
attempt to answer this question by considering propagating BICs on
2D periodic structures with reflection symmetries in
both periodic and orthogonal directions. A propagating BIC on a 2D
periodic structure is characterized by a real frequency and a real Bloch
wavenumber. Using a perturbation method, we show that when a small 
absorption is added to a periodic structure with a propagating BIC, 
a bound state emerges, but both the frequency and the Bloch
wavenumber of the bound state are complex. Our perturbation method
is also applicable to symmetry-protected BICs on periodic
structures. In that case, bound states with complex frequencies
emerge, but the Bloch wavenumber remains real and fixed. 

The rest of this paper is organized as follows. 
 The mathematical formulation and the definitions of various modes are
 recalled in 
 Sec.~II. The main steps of the perturbation theory are presented in
 Sec.~III. Numerical examples that validate the perturbation theory
 are shown in Sec.~IV. The paper is concluded with a summary and a
 brief discussion 
 in  Sec.~V.

\section{Formulation and definitions}

To analyze the effect of a small material loss on the BICs, we
consider  a 2D periodic structure that is invariant in $z$, 
periodic in $y$ with 
period $L$, bounded in the 
$x$ direction by $|x| < D$ for some $D>0$, and sandwiched by 
two homogeneous media given in $x>D$ and $x<-D$, respectively, 
where $\{x,y,z\}$ is a
Cartesian coordinate system. Let $\epsilon = \epsilon(x,y)$ be the
dielectric function describing the structure and the surrounding
media, then 
\begin{equation}
  \label{periodic}
\epsilon(x,y+L)=\epsilon(x,y) \quad \mbox{for all} \ (x,y),
\end{equation}
and
$\epsilon(x,y)$ is a positive constant for $x> D$ and $x<-D$,
respectively. For simplicity, we assume the surrounding medium is
vacuum, thus
\begin{equation}
  \label{vacuum}
  \epsilon(x,y)=1 \quad \mbox{for}\ |x| > D.   
\end{equation}

For  time-harmonic waves in the $E$ polarization, the $z$ component of
the electric field, denoted by $u$, satisfies the following Helmholtz
equation
\begin{equation}
\left[\partial^2_x+\partial^2_y+k^2\epsilon(x,y)\right]u=0,
\label{helm}
\end{equation}  
where  $k=\omega/c$ is the free space wavenumber, $\omega$ is the angular
 frequency, $c$ is the speed of light in vacuum, and the
time dependence is assumed to be $e^{-i \omega t}$. 
For a real frequency, an incident plane wave given in 
the homogeneous media ($x>D$  or $x< -D$) induces reflected
and transmitted  waves. If $(\alpha, \beta)$ is the wavevector of a
propagating plane wave, then $\alpha^2 + \beta^2 = k^2$, $\beta$ is
real, and $|\beta| < k$ so that $\alpha = \sqrt{k^2 - \beta^2}$ is also
real. Therefore, the radiation continuum for a 
fixed real $\beta$ is the frequency interval given by $\omega >
c|\beta|$. 

Without any incident field, electromagnetic waves 
may exist on the periodic structure as Bloch modes. 
For the 2D
structure and the $E$ polarization, a Bloch mode is a solution of
Eq.~\eqref{helm} given as 
\begin{equation}
\label{bloch}
 u(x,y) = e^{i\beta y}\phi(x,y),
\end{equation}
where $\phi$ is periodic in $y$ with the same period $L$, $\beta$ is
the Bloch wavenumber, and $u$ is
outgoing or exponentially decaying  as $|x| \to \infty$. The
periodicity of $\phi$ implies that the real part of $\beta$ can be
restricted by $ |\mbox{Re}(\beta) |   \le \pi/L$. 
For $|x|> D$, the
Bloch mode can be expanded in plane waves as 
\begin{equation}
  \label{pwaves}
u(x,y) = \sum_{j=-\infty}^\infty c_j^{\pm} e^{ i ( \beta_j y \pm
  \alpha_j x)}, \quad     \pm x > D,
\end{equation}
where $c_j^{\pm}$ are expansion coefficients, 
\begin{equation}
  \label{betaj}
\beta_j = \beta + \frac{2\pi j}{L}, \quad \alpha_j = \sqrt{ k^2 -
  \beta_j^2}. 
\end{equation}
The square root in
Eq.~\eqref{betaj} is chosen
such that either $\alpha_j$ has a positive imaginary part or
$\alpha_j$ is real and positive. 

If the periodic structure is a lossless dielectric 
structure, i.e., $\epsilon$ is real, it can have guided Bloch modes (with 
real $\beta$ and real $\omega$) that decay exponentially as $|x| \to 
\infty$. Typically, the guided modes exist below the lightline
(i.e. $ k < |\beta|$) and form bands that are continuous in $\beta$. 
In this case, all $\alpha_j$ are pure imaginary with positive
imaginary parts. A
BIC is a special guided mode above the lightline (i.e. $k >
|\beta|$). In that case, one or more $\alpha_j$ (including $\alpha_0 = \alpha$) are real and positive, but since the BIC decays exponentially as $|x|
\to \infty$, the corresponding coefficients $c_j^\pm$ must vanish. 
Clearly, a BIC does 
not couple with propagating plane waves with wavevectors $(\pm \alpha,
\beta)$.  
We call a BIC  a standing wave if $\beta=0$, and a propagating BIC if 
$\beta\ne 0$. Notice that the BICs only exist at special values of $\beta$.

On a 2D periodic structure, a resonant mode is a Bloch mode satisfying outgoing 
radiation conditions as $|x| \to \infty$ \cite{fan02}. It is usually defined for a
real $\beta$ and has a complex $\omega$. Since it radiates power to infinity,
the amplitude  of a resonant mode decays with time, and the $Q$-factor
can be defined as $Q = -0.5 \mbox{Re}(\omega)/\mbox{Im}(\omega)$. The
resonant modes form bands, and on each band they depend continuously on 
$\beta$, except for special values
of $\beta$ at which no power is radiated to infinity and $\omega$
becomes real. These special values of $\beta$ correspond exactly to
the BICs. Therefore, a BIC can be regarded as a special resonant mode
with an infinite $Q$-factor.  The resonant modes mostly exist 
above the lightline (defined using the real part of $k$), but their
endpoints are actually located below the lightline \cite{amgad19}. 

Leaky modes on wave-guiding structures are usually defined for a real
frequency \cite{marc91}.  For 2D periodic structures,  a leaky mode is also a Bloch
mode given in Eq.~\eqref{bloch}, it also
satisfies  outgoing radiation conditions as $|x| \to \infty$, but its
Bloch wavenumber $\beta$ is complex. If the leaky mode 
propagates toward $y=+\infty$, then $\mbox{Im}(\beta)$ should be
positive, such that as it propagates forward, it loses power and its
amplitude decays with $y$. Leaky modes can exist below and above the
lightline defined using the real part of $\beta$  \cite{amgad19}. 

For structures with a reflection symmetry in the periodic direction
(i.e., the $y$ direction), the antisymmetric standing waves (ASWs) are
well-known symmetry-protected BICs \cite{bonnet94,shipman03}. Assuming the origin is chosen such
that 
\begin{equation}
\label{eveniny}
\epsilon(x,y)=\epsilon(x,-y) \quad \mbox{for all} \ (x,y), 
\end{equation}
 an ASW is a special BIC with $\beta=0$ and an
antisymmetric wave field, i.e., $u(x,-y) = - u(x,y)$. 
Due to the symmetry mismatch between the wave field and the plane
waves with wavevectors $(\pm \alpha, \beta)  = (\pm k, 0)$, 
 the coefficients $c_0^\pm$ in Eq.~\eqref{pwaves} are automatically
 zero. The
 existence of ASWs can be rigorously proved \cite{bonnet94,shipman07},
 and they are robust under symmetric 
perturbations of the structure \cite{yuan17_4}. 

Periodic structures can have BICs that are not obviously
protected by any symmetry
\cite{port05,mari08,hsu13_1,bulg14b,gao16,yuan17,hu17josab,hsu13_2,yang14,zhen14,gan16,bulg15,bulg17}. In
the 2D case, it is relatively easy to find nontrivial
BICs on structures that are not only symmetric in $y$, but also
symmetric in $x$, i.e., 
\begin{equation}
\label{eveninboth}
\epsilon(x,y)=\epsilon(x,-y) = \epsilon(-x,y) \quad \mbox{for all} \ (x,y).
\end{equation}
If  $k<2\pi/L-|\beta|$, then all $\alpha_j$ for $j\ne 0$ are pure 
imaginary, thus, a BIC needs only to satisfy the condition $c_0^\pm = 0$. 
The reflection symmetry in $x$ implies that the BIC (with a
nonzero $\beta$ in general) is either even in $x$ or odd in
$x$. Therefore, the coefficients $c_0^+$ and $c_0^-$ are related by
$c_0^+ = \pm c_0^-$.  The reflection symmetry in $y$ is also
useful. It allows us to scale the BIC by a complex constant, such that 
\begin{equation}
\label{PTsym}
u(x,y) = \overline{u}(x,-y) \quad \mbox{for all}\ (x,y),
\end{equation} 
where $\overline{u}$ is the complex conjugate of
$u$ \cite{port05,yuan17_4}. This is a case of the $\mathcal{PT}$-symmetry. It is easy to
verify that the periodic function $\phi$ in Eq.~\eqref{bloch} is also
$\mathcal{PT}$-symmetric. 

\section{Perturbation theory}

Suppose there is a BIC on a lossless dielectric structure described by a real dielectric
function $\epsilon$, it is clearly important to find out what 
happens to the BIC, if the structure is slightly perturbed \cite{yuan17_4,kosh18}. In
general, a real perturbation of the dielectric function $\epsilon$
will destroy the BIC and produce a resonant 
mode with a complex frequency, but special perturbations may preserve
the BIC \cite{yuan17_4}. The case for symmetry-protected  BICs is well understood. 
If the dielectric function $\tilde{\epsilon}$ of the perturbed structure is
still real and symmetric, the BIC is preserved and will have a
slightly different real frequency. The case for BICs without 
symmetry protection (i.e., there is no symmetry mismatch between the
BIC and the radiating waves) is more complicated. For some cases, it is still
possible to identify special perturbations that preserve the 
BIC. For example, a propagating BIC on a periodic structure with
reflection symmetry in both $x$ and  
$y$ directions is preserved if the perturbation is also symmetric in
$x$ and $y$ \cite{yuan17_4}. The perturbed structure will have a BIC with a
slightly different frequency  and a slightly different Bloch
wavenumber. 

Here, we investigate the effect of a small material loss on the
BICs. The dielectric function of a lossy periodic structure is
written as 
\begin{equation}
\label{lossy}
\tilde{\epsilon}(x,y) = \epsilon(x,y)+\delta F(x,y),
\end{equation} 
where $\epsilon$ is real,  $F$
is a complex function with $\mbox{Im}(F) \ge 0$ and
$\max [\mbox{Im}(F)] = 1$, 
and $\delta$ is a small positive number.  We study the
problem by a perturbation method, assuming the lossless structure
given by $\epsilon(x,y)$ has a BIC $u =\phi \exp(  i \beta y)$
with frequency $\omega$, and the lossy structure has a Bloch mode 
$\tilde{u} =\tilde{\phi} \exp( i \tilde{\beta} y)$ with frequency $\tilde{\omega}$. 
Due to the material loss, it is clearly impossible to have a real
$\tilde{\omega}$ and a real $\tilde{\beta}$, if the Bloch mode
$\tilde{u}$ is required to decay exponentially or radiate power outward
as $|x| \to \infty$. Moreover, if $\tilde{u}$ is allowed to radiate
power to infinity, we can specify a real $\tilde{\beta}$ around $\beta$
and find a Bloch mode with a complex $\tilde{\omega}$, or specify a real 
$\tilde{\omega}$ around $\omega$ and calculate a solution with a complex 
$\tilde{\beta}$. These solutions correspond to resonant
and leaky modes, respectively, and they are not the topic of this
paper. Since the BIC decays exponentially as $|x| \to \infty$, we look
for a perturbed mode $\tilde{u}$ that also decays exponentially as $|x| \to
\infty$. More precisely, we assume the frequency and Bloch wavenumber of
the original BIC satisfy 
\begin{equation}
  \label{onerad}
|\beta| < k < \frac{2\pi}{L} - |\beta|, 
\end{equation}
and look for a nearby bound state $\tilde{u}$ such that the coefficients $c_0^\pm$ for
$\tilde{u}$ (defined as in Eq.~\eqref{pwaves}) vanish. For that
purpose, we need to determine both $\tilde{\omega}$ and
$\tilde{\beta}$. 

Let us consider a lossless periodic structure with a real dielectric function $\epsilon$
satisfying Eqs.~\eqref{periodic}, \eqref{vacuum} and
\eqref{eveninboth}, and a BIC with frequency $\omega$ and a Bloch
wavenumber 
$\beta$ satisfying Eq.~\eqref{onerad}, and assume the BIC is scaled to
satisfy the ${\cal PT}$-symmetric 
condition \eqref{PTsym}. If $\tilde{u}$ is a bound state (close to the
BIC) on the lossy structure with a dielectric function $\tilde{\epsilon}$ given in
Eq.~\eqref{lossy}, then $\tilde{u}$ satisfies Helmholtz equation
\eqref{helm} with $\epsilon$ and $k$ replaced by $\tilde{\epsilon}$
and $\tilde{k} = \tilde{\omega}/c$, respectively. To find $\tilde{u}$, 
we use a perturbation method by expanding $\tilde{\phi}$, 
$\tilde{k}$  and $\tilde{\beta}$ in power series of
$\delta$ \cite{yuan17_4}: 
\begin{eqnarray}
  \label{powerindel}
\tilde{\phi} &=&\phi+\delta\phi_1+\delta^2\phi_2+\cdots   \\
\tilde{\beta} &=& \beta +\delta\beta_1+\delta^2\beta_2+\cdots \\
\tilde{k} &=& k +\delta k_1+\delta^2 k_2+\cdots 
\end{eqnarray}
Inserting the above into the governing equation of $\tilde{\phi}$
(derived from the Helmholtz equation for $\tilde{u}$, see Appendix), and considering
the different powers of $\delta$, we obtain
${\cal L} \phi = 0$ for 
\begin{equation}
\label{opL}
\mathcal{L}
= \partial^2_x+\partial^2_y+2i\beta\partial_y+k^2\epsilon 
      -\beta^2, 
\end{equation}
and 
\begin{equation}
\label{eqphij}
  {\cal L} \phi_j  = B_1\beta_j+B_2k_j-C_j, 
\end{equation}
for $j \ge 1$,   where 
\[
  B_1 = 2\beta\phi-2i\partial_y \phi,  \quad
  B_2  =  -2k\epsilon\phi, \quad 
  C_1 = k^2F\phi,
\]
and $C_j$ for $j\ge 2$ are given in Appendix. 

The operator ${\cal L}$ is singular, since ${\cal L} \phi = 0$ has a
nonzero solution corresponds to the BIC. Equation~\eqref{eqphij} is a Helmholtz equation with a source term in the right
hand side. It does not
have a solution, unless the right hand side is orthogonal with the
BIC, that is
\begin{equation}
  \label{orthbic}
\int_\Omega \overline{\phi} \left( B_1\beta_j+B_2k_j-C_j \right) d{\bf  r} = 0,
\end{equation}
where ${\bf r}=(x,y)$ and $\Omega$ is the domain for one period of the
structure, given by 
\[
\Omega  = \{  (x,y) \ | \ -\infty < x < \infty, \ -L/2 <
y < L/2 \}. 
\]
Since $\tilde{u}$ is required to have $c_0^\pm = 0$ (defined as in 
Eq.~\eqref{pwaves}), $\phi_j$ must decay exponentially as $|x| \to
\infty$, then the right hand side of Eq.~\eqref{eqphij} must be
orthogonal to a diffraction solution $\varphi$ for the corresponding
frequency $\omega$ and wavenumber $\beta$. That is, 
\begin{equation}
\label{orthdif}
\int_{\Omega}\overline{\varphi} \left( B_1\beta_j+B_2k_j-C_j 
\right) d {\bf r}=0. 
\end{equation}
More details on $\varphi$ are given in Appendix. In particular,
$\varphi$ is chosen to be even (or odd) in $x$ if the BIC is 
even (or odd) in $x$, and it is scaled to satisfy the ${\cal
  PT}$-symmetry condition \eqref{PTsym}. 
Equations~\eqref{orthbic} and \eqref{orthdif} give rise to 
\begin{equation}
\label{bus}
{\bf A} \left[
\begin{matrix} \beta_j \cr k_j \end{matrix}
\right] 
=\left[\begin{matrix} b_{1j} \cr  b_{2j} \end{matrix}\right], 
\quad 
{\bf A} = \left[\begin{matrix} a_{11} & a_{12} \cr  a_{21} & a_{22} \end{matrix}\right]
\end{equation}
where
\begin{eqnarray*}
& a_{11}=\int_{\Omega}\overline{\phi} B_1d\bf{r},  
& \hspace{1cm} 
a_{12} = \int_{\Omega}\overline{\phi} B_2d\bf{r}, \\
& a_{21} = \int_{\Omega}\overline{\varphi} B_1d\bf{r},
& \hspace{1cm}  a_{22} = \int_{\Omega}\overline{\varphi} B_2d\bf{r}, \\
& b_{1j} = \int_{\Omega}\overline{\phi} C_jd\bf{r},
& \hspace{1cm}  b_{2j} = \int_{\Omega}\overline{\varphi} C_jd\bf{r}.
\end{eqnarray*}

Since $\phi$ and $\varphi$ satisfy the ${\cal PT}$-symmetry condition
\eqref{PTsym}, and $\epsilon$ satisfies Eq.~\eqref{eveninboth}, it is
easy to show that all entries of matrix ${\bf A}$ are real. If ${\bf
  A}$ is invertible, we can solve $\beta_j$ and $k_j$ from the above
system, and then solve $\phi_j$ from Eq.~\eqref{eqphij}. The two
conditions \eqref{orthbic} and \eqref{orthdif} ensure that
Eq.~\eqref{eqphij} has a solution that decays exponentially as $|x|
\to \infty$. Moreover, $\phi_j$ has the same parity (even or odd in
$x$) as the BIC. 
For the lossy structure, the function $F$ in Eq.~\eqref{lossy} is
assumed to be pure imaginary. If we assume that $F$ is also
symmetric in both $x$ and $y$, i.e., $F$ satisfies
Eq.~\eqref{eveninboth}, then $b_{11}$ and $b_{21}$ are also pure
imaginary. This means that $\beta_1$ and $k_1$ are pure imaginary and
nonzero in general.  Therefore, both $\tilde{\omega}$ and
$\tilde{\beta}$ of the bound state $\tilde{u}$ are complex in
general. 

The ASWs are important special cases with $\beta = 0$. The
field profile $\phi$ of an ASW is odd in $y$. The diffraction solution
$\varphi$ is even in $y$. To be consistent with the  
above perturbation theory, we can scale $\phi$ 
as a pure imaginary function, and  scale 
$\varphi$ as a real function, then both satisfy 
Eq.~\eqref{PTsym}. Due to these properties, it is easy to show that
$a_{22} = 0$ and $b_{21}=0$.  It is also clear that 
$a_{12} \ne 0$. If $a_{21} \ne 0$, we 
conclude that $\beta_1=0$ and  $k_1 = b_{11}/a_{12}$.  Similar to the general case above,
Eq.~\eqref{eqphij} can be solved, and  the solution 
$\phi_j$ is also odd in $y$. This implies that $b_{2j}=0$ and $\beta_j =
0$ for all $j \ge 1$.  Furthermore, for the ASWs, the reflection
symmetry in $x$ is not important. We can obtain the same conclusion
if $\epsilon$ and $F$ have only a reflection symmetry in $y$, that is, they satisfy
Eq.~\eqref{eveniny}, instead of Eq.~\eqref{eveninboth}.

 The perturbation theory is valid only when the matrix ${\bf A}$ is
 invertible or $a_{21} \ne 0$ if the BIC is an ASW. The matrix ${\bf
   A}$ can indeed be singular, for example, when the BIC is a symmetric
 standing wave, i.e., $\beta = 0$ and $\phi$ is even in $y$. A similar
 restriction is present in the theory on the robustness of the
 BICs \cite{yuan17_4}. However, for a generic BIC, the matrix ${\bf
   A}$ is typically invertible.

\section{Numerical examples}

In this section, we present some numerical results to validate the
perturbation theory. The periodic structure concerned is an  array of 
parallel, identical and infinitely long circular cylinders with radius
$a$ and dielectric constant $\epsilon_c$ surrounded by vacuum.
The cylinders are parallel to the $z$ axis and their centers are
located on the $y$ axis. The period in the $y$ direction, i.e. the
center-to-center distance between two nearby cylinders,  is $L$. 
The structure is chosen for its
simplicity. In fact, a semi-analytic computational method based on cylindrical
and plane wave expansions is available to calculate Bloch modes of the
structure to high accuracy \cite{hu15}. 

\begin{figure}[htbp]
\centering
\includegraphics[width=\linewidth]{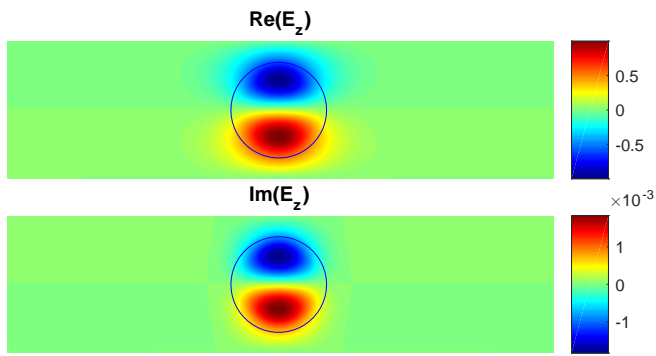}
\caption{Electric field pattern of a bound state near an ASW for
  $\delta=0.1$.} 
\label{sbice}
\end{figure}

Antisymmetric standing waves  on periodic arrays of circular cylinders
have been thoroughly investigated before
\cite{shipman03,bulg14b,hu15}. On an array of cylinders with 
$a=0.3L$ and $\epsilon_c=11.6$, there is an ASW with normalized frequency 
$kL/(2\pi)=0.41122783$. Now we assume the cylinders are lossy, with a
dielectric constant $\epsilon_c= 11.6 + i\delta$ for $\delta>
0$.  According to the perturbation theory, when $\delta$ is small,
there should be a bound state with a complex frequency (and a zero
Bloch wavenumber) near the original ASW. 
We calculate the bound state for a few different values of $\delta$
and list the  numerical results for $\tilde{k}$ in 
Table~\ref{sbic}
\begin{table}[htbp]
\centering 
\caption{Complex $\tilde{k}$ of the bound state near an ASW for
  a few values of $\delta$.}
\begin{tabular}{lc}
\hline 
$\delta$ & $\tilde{k} L$ \\
\hline 
0.1  & 2.5837547 - 0.01068485i       \\
0.01 & 2.5838200 - 0.00106853i       \\
0.001 & 2.5838207 - 0.00010685i     \\
0.0001  & 2.5838207 - 0.00001069i   \\
\hline 
\end{tabular}
  \label{sbic}
\end{table}
below. The field pattern of the bound state for $\delta = 0.1$ is 
shown in Fig.~\ref{sbice}.
As we know from Sec.~III, the bound state near the ASW is also
anti-symmetric (odd in $y$) and also decays exponentially as $x \to \pm
\infty$. These properties are noticeable in Fig.~\ref{sbice}. 
The perturbation theory gives us the analytic formula 
$k_1=b_{11}/a_{12}$ for the first order term in the power series of
$\tilde{k}$,  where $b_{11}$ and $a_{12}$ are related to the ASW, and functions $\epsilon$ and 
$F$. Since all these are known, we evaluate $k_1$ by this formula and obtain  
$k_1L  \approx -0.1068i$. From the numerical solutions of $\tilde{k}$
for different values of $\delta$, we can also estimate $k_1 = \partial
\tilde{k}/\partial \delta|_{\delta = 0}$ by a difference
formula. The numerical result is 
$k_1 L \approx  -0.1069i$. The agreement with the analytic formula is
excellent.

\begin{figure}[htbp]
\centering
\includegraphics[width=\linewidth]{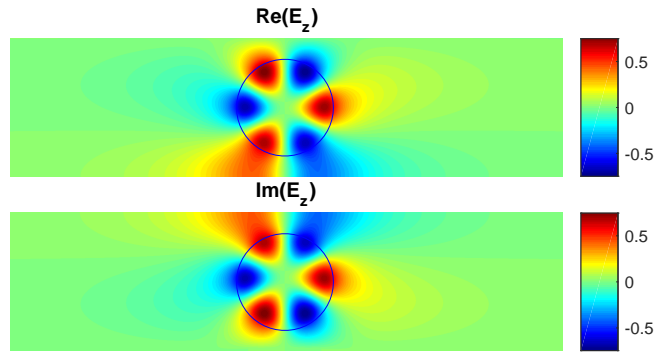}
\caption{Electric field pattern of a bound state near an odd-in-$x$
  propagating BIC for $\delta=0.1$.}
\label{pbicodd}
\end{figure}

A periodic array of circular cylinders can support propagating
BICs \cite{bulg14b,yuan17}. We assume the cylinders have a radius $a =
0.35L$ and dielectric constant $\epsilon_c  =11.56 + i \delta$. For
$\delta=0$, the periodic array has  
one odd-in-$x$ propagating BIC with $kL/(2\pi) = 0.6702$ and
$\beta/(2\pi)=0.2483$, and one even-in-$x$ propagating BIC with
$kL/(2\pi)=0.4854$ and $\beta L/(2\pi)=0.0780$. For a small $\delta >
0$, these two BICs are turned to bound states with a complex 
frequency and a complex Bloch wavenumber. We calculate the bound states
for a few different values of $\delta$. The numerical results for 
$\tilde{k}$ and $\tilde{\beta}$ of the bound state near the odd-in-$x$
BIC are listed in Table~\ref{pbico}. 
\begin{table}[htbp]
\centering
\caption{Complex $\tilde{k}$ and $\tilde{\beta}$ of a
  bound state near an odd-in-$x$ propagating BIC for a few values
  of $\delta$.}
\begin{tabular}{lcc}
\hline
$\delta$ & $\tilde{k} L $ & $\tilde{\beta}L$\\
\hline
0.1  & 4.2111340 - 0.01732564i & 1.5603202 - 0.00621566i    \\
0.01 & 4.2112313 - 0.00173248i   & 1.5603429 - 0.00061963i   \\
0.001 & 4.2112322 - 0.00017322i  &  1.5603422 - 0.00006146i \\
0.0001  & 4.2112323 - 0.00001733i  &  1.5603430 - 0.00000633i \\
\hline
\end{tabular}
  \label{pbico}
\end{table}
The field pattern of the bound state for $\delta=0.1$ 
is shown in Fig.~\ref{pbicodd}.
The perturbation theory allows us to solve $\beta_j$ and $k_j$ from
the $2\times 2$ linear system Eq.~\eqref{bus}. The first order terms
$\beta_1$ and $k_1$ are particularly easy to evaluate, since they are
only related to the BIC, the diffraction solution $\varphi$, and the
given functions $\epsilon$ and $F$. Using the odd-in-$x$ solution
$\varphi$, we obtain $\beta_1 L \approx -0.0623i$ and 
$k_1 L \approx -0.1732i$. Since $k_1$ and $\beta_1$ are partial
derivatives of $\tilde{k}$ and $\tilde{\beta}$ with respect to
$\delta$ and evaluated at $\delta=0$, we can estimate $k_1$ and
$\beta_1$ from the numerical solutions in Table~\ref{pbico} by
some difference formulae. The estimated values are 
$\beta_1 L \approx -0.0620i$ and $k_1L \approx -0.1733i$ and they
agree very well with the exact values by the perturbation theory.

\begin{figure}[htbp]
\centering
\includegraphics[width=\linewidth]{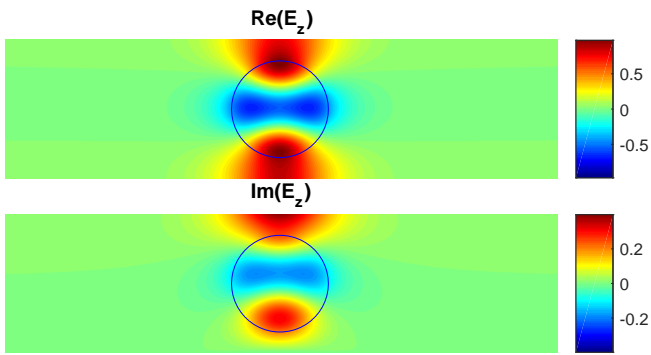}
\caption{Electric field pattern of a bound state near an even-in-$x$
  propagating BIC for $\delta=0.1$.}
\label{pbiceven}
\end{figure}

For the same periodic array, we calculate the bound state
near the even-in-$x$ propagating BIC.
The complex $\tilde{k}$ and $\tilde{\beta}$ of the bound state are
listed in Table~\ref{pbice}. 
\begin{table}[htbp]
\centering
\caption{Complex $\tilde{k}$ and $\tilde{\beta}$ of a bound state near
  an even-in-$x$ propagating BIC for a few values of $\delta$.}
\begin{tabular}{lcc}
\hline
$\delta$ & $\tilde{k}L$ & $\tilde{\beta} L$\\
\hline
0.1  & 3.0498207 - 0.01286425i & 0.49006788 - 0.00786603i    \\
0.01 & 3.0499008 - 0.00128612i   & 0.49006724 - 0.00078319i   \\
0.001 & 3.0499016 - 0.00012832i  &  0.49006700 - 0.00007557i \\
0.0001  & 3.0499017 - 0.00001262i  &  0.49006845 - 0.00000555i \\
\hline
\end{tabular}
  \label{pbice}
\end{table}
The field pattern of the bound state for $\delta=0.1$ 
is shown in Fig.~\ref{pbiceven}.
Using an even-in-$x$ diffraction solution, we evaluate 
$\beta_1$ and $k_1$ by solving Eq.~\eqref{bus}, and obtain 
$\beta_1 L \approx -0.0786i$  and $k_1 L \approx -0.1286i$. Up to the
first four significant digits, the
estimated values based on the numerical solutions given in Table~\ref{pbice}
are exactly the same.

\section{Conclusion}

An optical BIC usually exists on an open lossless
dielectric structure. It has a real frequency, and also a real
Bloch wavevector if the structure is periodic. We study how BICs are
affected when a small material loss is added to the structure. The result
on symmetry-protected BICs is fully expected. For 2D periodic
structures with a reflection symmetry, a symmetry-protected BIC
becomes a bound state with a complex frequency, 
if the absorption profile preserves the relevant symmetry. The new bound
state still has a symmetry 
mismatch with the radiative waves, and does not radiate power to
infinity. The imaginary part of the frequency is solely the
consequence of material loss, and the real part of the frequency remains in the
radiation continuum. 

The case for BICs without the usual symmetry-protection is more
complicated. In general, adding a small material loss to the structure may
open up a radiation channel and turn a BIC to a resonant mode that
radiates power to infinity. Many BICs are propagating modes with a real
and nonzero Bloch wavevector on periodic structures with certain
symmetry, and their frequencies are within the interval where only
one radiation channel exists. These propagating BICs are not
protected by symmetry in the 
sense of a symmetry mismatch, but they depend on the symmetry for
their existence and robustness \cite{yuan17_4}. We consider 2D periodic structures with
reflection symmetry in both periodic and perpendicular directions,
and study the effect of a small material loss that preserves the
reflection symmetry. We show that 
a generic propagating BIC is turned to a bound state with a complex
frequency and a complex Bloch wavenumber. The real part of the complex
frequency still lies in the radiation continuum. The 
imaginary part accounts for both radiation and absorption losses. 
The complex Bloch
wavenumber implies that the bound state no longer has a uniform
magnitude along the periodic direction. 
In general, it is impossible to have a bound state with a real Bloch wavenumber
near the original propagating BIC. All modes with a real Bloch
wavenumber at and near that of the BIC are resonant modes. 


The performance of devices based on BICs may be limited by a number of
practical issues, such as material losses, fabrication errors, finite sizes (as true BICs
only exist on infinite structures), and loss of periodicity (for
periodic structures). The impact of material loss to resonant effects
($Q$-factors and field enhancement) around a BIC  has been analyzed by
Yoon {\it et al.} \cite{yoon15}. Our study on bound states with
complex frequencies reveals a major difference between BICs protected
or unprotected by symmetry. It is worthwhile to investigate the
consequences of this difference on the resonant effects.

\section*{Appendix}

The perturbation theory follows our earlier work \cite{yuan17_4}. Some key
steps are listed below. More details can be found in \cite{yuan17_4}. 
The governing equation for $\tilde{\phi}$ is
\[
\left[ \partial_x^2 + \partial_y^2 + \tilde{k}^2 \tilde{\epsilon} -
\tilde{\beta}^2 + 2 i \tilde{\beta} \partial_y \right] \tilde{\phi} = 0.
\]
In the right hand side of Eq.~\eqref{eqphij}, $C_j$ (for $j\ge 2$) are
given by 
\[
\begin{array}{l}
C_j =\left[\sum\limits_{l=1}^{j-1}\left(k_lk_{j-l}\epsilon)-\beta_l\beta_{j-l}\right) 
+F\sum\limits_{l=0}^{j-1}k_lk_{j-1-l}\right]\phi \\
+\sum\limits_{n=1}^{j-1}\left[\sum\limits_{l=0}^{n}\left(k_lk_{n-l}\epsilon-\beta_l\beta_{n-l}\right) 
+F\sum\limits_{l=0}^{n-1}k_lk_{n-1-l}\right]\phi_{j-n} \\
+2i\sum\limits_{n=1}^{j-1}\beta_n\partial_y\phi_{j-n}
\end{array}
\]
where $\beta_0=\beta$ and $k_0=k$. 

Equation~\eqref{orthbic} is derived by multiplying $\overline{\phi}$
to both sides of Eq.~\eqref{eqphij}, integrating on $\Omega$,
and showing that  
\[
\int_\Omega \overline{\phi} {\cal L} \phi_j d{\rm r}= 0.
\]

If the BIC is even in $x$, we consider a diffraction problem with
two incident plane waves:  $\exp[i (\beta y + \alpha x)]$ for $x < -D$
and $\exp[i (\beta y - \alpha x)]$ for $x > D$. Since the structure is
symmetric in $x$, the diffraction solution, denoted as $\tilde{v}_e$,
is also even in $x$. Moreover, $\tilde{v}_e$ contains outgoing plane
waves of identical amplitudes as $x \to \pm \infty$, that is
\[
\tilde{v}_e(x,y) \sim e^{ i (\beta y \pm \alpha x)} + S_e e^{i (\beta
  y \mp \alpha x)}, \quad x \to \mp \infty.
\]
Since $\epsilon$ is real, energy is conserved, thus $|S_e|=1$. If $S_e
= \exp( 2i \theta_e)$ for a real $\theta_e$, we define 
$v_e = \tilde{v}_e \exp(-i \theta_e)$, then $v_e$ is even in $x$ and
satisfies the ${\cal PT}$-symmetry condition, Eq.~\eqref{PTsym}. The
function $\varphi$ is given by $\varphi = v_e \exp(-i \beta y)$. If
the BIC is odd in $x$, we similarly define a diffraction solution
$v_o$ which is odd in $x$ and satisfies Eq.~\eqref{PTsym}, and let 
$\varphi = v_o \exp(-i \beta y)$.

Equation~\eqref{orthdif} is derived by multiplying $\overline{\varphi}$ to both
sides of Eq.~\eqref{eqphij} and integrating on the rectangle
$\Omega_h$ for $|x| < h$ and $|y| < L/2$. 
Assuming $\phi_j \to 0$ as $|x| \to \infty$
exponentially and taking the limit for $h \to \infty$, 
the left hand side gives 
\[
\int_\Omega \overline{\varphi} {\cal L} \phi_j d{\rm r}= 0.
\]
The right hand side must also vanish. This leads to
Eq.~\eqref{orthdif}. On the other hand, if $\beta_j$ and $k_j$
satisfy Eq.~\eqref{bus}, then Eq.~\eqref{eqphij} (for $\phi_j$)  has
a solution and that solution must decay exponentially as $|x| \to
\infty$. 

\section*{Acknowledgments}
The authors acknowledge support from
the Fundamental Research Funds for the Central 
Universities of China (Grant No. 2018B19514), 
 the Science and Technology
Research Program of Chongqing Municipal Education Commission, China
(Grant No. KJ1706155), and 
the Research Grants Council of
Hong Kong Special Administrative Region, China (Grant No. CityU
11304117).


\begin{thebibliography}{99}

\bibitem{hsu16} C. W. Hsu, B. Zhen, A. D. Stone,
  J. D. Joannopoulos, and M. Solja\v{c}i\'{c}, 
 ``Bound states in the continuum,'' 
Nat. Rev. Mater. {\bf 1}, 16048 (2016).

\bibitem{kosh19} K. Koshelev, G. Favraud, A. Bogdanov, Y. Kivshar, 
and A. Fratalocchi, ``Nonradiating photonics with resonant dielectric
nanostructures,'' Nanophotonics {\bf 8}, 725--745 (2019). 

\bibitem{neumann29}  J. von Neumann and E. Wigner, 
 ``\"{U}ber   merkw\"{u}rdige diskrete eigenwerte,'' 
Z. Physik  {\bf 50},  291-293 (1929). 

\bibitem{urse87} F. Ursell, 
  ``Mathematical aspects of trapping modes in the theory of surface waves,''
  J . Fluid Mech. {\bf 183}, 421-437 (1987).

 \bibitem{evans91} D. V. Evans and C. M. Linton, 
  ``Trapped modes in open channels,'' 
 J. Fluid Mech. {\bf 225}, 153-175 (1991). 



 \bibitem{evans94}  D. V. Evans, M. Levitin and  D. Vassiliev,
   ``Existence theorems for trapped modes,''
 J. Fluid Mech. {\bf 261}, 21-31 (1994).




\bibitem{longhi07} 
S. Longhi, ``Bound states in the continuum in a single-level
Fano-Anderson model,'' Eur. Phys. J. B  {\bf 57}, 45--51 (2007). 

\bibitem{plot11} Y. Plotnik, O. Peleg, F. Dreisow, M. Heinrich,
  S. Nolte, A. Szameit, and M. Segev, 
 ``Experimental observation of optical bound states in the
 continuum,'' 
\prl\ {\bf 107}, 183901 (2011).
 


\bibitem{weim13} S. Weimann, Y. Xu, R. Keil, A. E. Miroshnichenko,
A. T\"{u}nnermann, S. Nolte, A. A. Sukhorukov, A. Szameit, and Y. S. Kivshar,
 ``Compact surface Fano states embedded in the continuum of the
 waveguide arrays,''  
\prl\ {\bf 111}, 240403 (2013). 

\bibitem{zou15} C. L. Zou, J.-M. Cui, F.-W. Sun, X. Xiong, X.-B. Zou,
   Z.-F. Han, and G.-C. Guo, 
  ``Guiding light through optical bound states in the continuum for
  ultrahigh-Q microresonantors,''  
   Laser \& Photonics Rev.  {\bf 9}, 114-119 (2015). 

\bibitem{shipman13} S. P. Shipman and A. T. Welters, ``Resonant
  electromagnetic scattering in anisotropic layered media,''
  J. Math. Phys.  {\bf 54}, 103511 (2013).

\bibitem{gomis17} J. Gomis-Bresco, D. Artigas,  and L. Torner, 
  ``Anisotropy-induced photonic bound states in the continuum,''
  Nature Photonics  {\bf 11}, 232--237 (2017). 


 \bibitem{bonnet94} A.-S. Bonnet-Bendhia and F. Starling, ``Guided
 waves by electromagnetic gratings and nonuniqueness examples for the
 diffraction problem,''  Math. Methods Appl. Sci.  {\bf 17}, 305-338
 (1994).  

\bibitem{shipman03} S. P. Shipman and S. Venakides, 
``Resonance and bound states in photonic crystal slabs,'' 
SIAM J. Appl. Math.  {\bf 64}, 322-342 (2003). 

\bibitem{port05} R. Porter and D. Evans, ``Embedded Rayleigh-Bloch 
  surface waves along periodic rectangular arrays,''  Wave Motion 
  {\bf 43}, 29-50 (2005). 

\bibitem{shipman07} S. Shipman and D. Volkov, ``Guided modes in 
  periodic slabs: existence and nonexistence,''  
SIAM J. Appl. Math. {\bf 67}, 687--713 (2007). 

\bibitem{mari08} D. C. Marinica, A. G. Borisov, and 
  S. V. Shabanov, ``Bound states in the continuum in photonics,'' 
  \prl\ {\bf 100}, 183902 (2008).   


 \bibitem{hsu13_1} C. W. Hsu, B. Zhen, S.-L. Chua, S. G. Johnson,
   J. D. Joannopoulos, and M. Solja\v{c}i\'{c}, ``Bloch surface 
   eigenstates within the radiation continuum,'' 
  Light Sci. Appl. {\bf 2}, e84 (2013). 

\bibitem{bulg14b} E. N. Bulgakov and A. F. Sadreev, ``Bloch 
  bound states in the radiation continuum in a periodic array of 
  dielectric rods,''   \pra\ {\bf 90}, 053801 (2014). 

\bibitem{hu15} Z. Hu and Y. Y. Lu, ``Standing waves on two-dimensional 
  periodic dielectric waveguides,''  Journal of Optics {\bf 17},
  065601 (2015).  

\bibitem{gao16} X. Gao, C. W. Hsu, B. Zhen, X. Lin,
  J. D. Joannopoulos, M. Solja\v{c}i\'{c}, and H. Chen, ``Formation mechanism 
  of guided resonances and bound states in the continuum in photonic 
  crystal slabs,'' Sci. Rep. {\bf 6}, 31908 (2016). 

\bibitem{yuan17} L. Yuan and Y. Y. Lu, ``Propagating Bloch modes 
  above the lightline on a periodic array of cylinders,'' 
  J. Phys. B: Atomic, Mol. and Opt. Phys. {\bf 50}, 05LT01 (2017). 

 \bibitem{hu17josab} Z. Hu and Y. Y. Lu, ``Propagating 
     bound states in the continuum at the surface of a photonic 
     crystal,'' \josab\ {\bf 34}, 1878-1883 (2017). 


\bibitem{yuan17_4} L. Yuan and Y. Y. Lu,  ``Bound states in the 
  continuum on periodic structures: perturbation theory and 
  robustness,'' \ol\ {\bf 42}(21) 4490-4493 (2017). 

\bibitem{bulg17pra} E. N. Bulgakov and D. N. Maksimov, 
``Bound states in the continuum and polarization singularities in 
periodic arrays of dielectric rods,'' 
\pra\ {\bf 96}, 063833 (2017). 

\bibitem{hu18} Z. Hu and Y. Y. Lu, ``Resonances and bound states in 
  the continuum on periodic arrays of slightly noncircular 
  cylinders,'' J. Phys. B: At. Mol. Opt. Phys.  {\bf 51}, 035402 (2018). 

\bibitem{doel18} H. M. Doeleman, F. Monticone, W. den Hollander,
  A. Al\`{u}, and A. F. Koenderink,
``Experimental observation of a polarization vortex at an optical
bound state in the continuum,'' Nature Photonics {\bf 12}, 397--401
(2018). 


\bibitem{padd00} P. Paddon and J. F. Young, ``Two-dimensional 
  vector-coupled-mode theory for textured planar waveguides,'' 
\prb\ {\bf 61}, 2090-2101 (2000). 

\bibitem{tikh02} S. G. Tikhodeev, A. L. Yablonskii, E. A Muljarov,
  N. A. Gippius, and T. Ishihara, 
``Quasi-guided modes and optical properties of photonic crystal 
slabs,'' \prb\ {\bf 66}, 045102 (2002). 

\bibitem{lee12} J. Lee, B. Zhen, S. L. Chua, W. Qiu, J. D. Joannopoulos,
  M. Solja\v{c}i\'{c}, and O. Shapira, ``Observation and 
  differentiation of unique high-Q optical resonances near zero wave 
  vector in macroscopic photonic crystal slabs,'' \prl\ {\bf 109},
  067401 (2012). 

\bibitem{hsu13_2} C. W. Hsu, B. Zhen, J. Lee, S.-L. Chua,
  S. G. Johnson, J. D. Joannopoulos, and M. Solja\v{c}i\'{c},
  ``Observation of trapped light within the radiation continuum,'' 
  Nature {\bf 499}, 188--191 (2013). 

\bibitem{yang14} Y. Yang, C. Peng, Y. Liang, Z. Li, and S. Noda, ``Analytical 
perspective for bound states in the continuum in 
photonic crystal slabs,'' \prl\ {\bf 113}, 037401 (2014). 

\bibitem{zhen14} B. Zhen, C. W. Hsu, L. Lu, A. D. Stone, and M. 
Solja\v{c}i\v{c},  ``Topological nature of optical bound 
states in the continuum,'' 
\prl\ {\bf 113}, 257401 (2014). 

\bibitem{gan16} R. Gansch, S. Kalchmair,  P. Genevet,
  T. Zederbauer, H. Detz,  A. M. Andrews, W. Schrenk, F. Capasso,  M. Lon\v{c}ar, and 
 G. Strasser, ``Measurement of bound states in the continuum by a 
 detector embedded in a photonic crystal,'' Light: Science \&
 Applications {\bf 5}, e16147 (2016). 

\bibitem{li16} L. Li and H. Yin, ``Bound states in the continuum in 
double layer structures,'' Sci. Rep. {\bf 6}, 26988 (2016). 

\bibitem{ni16} L. Ni, Z. Wang, C. Peng, and Z. Li, ``Tunable optical 
  bound states in the continuum beyond in-plane symmetry protection,'' 
\prb\ {\bf 94}, 245148  (2016). 

\bibitem{zhang18} Y. Zhang, A. Chen, W. Liu, C. W. Hsu, B. Wang, F. Guan,
  X. Liu, L. Shi, L. Lu, and J. Zi, 
  ``Observation of polarization vortices in momentum space,''
  \prl\  {\bf 120}, 186103 (2018). 


 \bibitem{bulg15} E. N. Bulgakov and A. F. Sadreev, ``Light trapping 
  above the light cone in one-dimensional array of dielectric 
  spheres,'' 
 \pra\ {\bf 92}, 023816 (2015). 


\bibitem{bulg17} E. N. Bulgakov and A. F. Sadreev, ``Bound states in 
  the continuum with high orbital angular momentum in a dielectric 
 rod with periodically modulated permittivity,'' \pra\ {\bf 96}, 013841 (2017). 



\bibitem{kosh18} K. Koshelev, S. Lepeshov, M. Liu, A. Bogdanov, and
  Y. Kivshar, ``Asymmetric metasurfaces with high-$Q$ resonances
  governed by bound states in the continuum,''
  \prl\  {\bf 121}, 193903 (2018). 

\bibitem{lijun17pra} L.  Yuan and Y. Y. Lu,
``Strong resonances on periodic arrays of cylinders and optical
bistability with weak incident waves,'' 
\pra\ {\bf 95}, 023834 (2017).

\bibitem{lijun18pra} L.  Yuan and  Y. Y.  Lu, 
  ``Bound states in the continuum on periodic structures surrounded by
  strong resonances,'' 
\pra\ {\bf 97}, 043828 (2018). 


\bibitem{kodi17} A. Kodigala, T. Lepetit, Q. Gu, B. Bahari,
  Y. Fainman, and B. Kant\'{e}, ``Lasing action from photonic bound 
  states in continuum,'' Nature {\bf 541}, 196-199 (2017).

\bibitem{romano19} S. Romano, A. Lamberti, M. Masullo, E. Penzo, 
  S. Cabrini, I. Rendina, and V. Mocella, 
  ``Optical biosensors based on photonic crystals
supporting bound states in the continuum,'' Materials {\bf 11}, 526 (2018). 

\bibitem{yesi19} F. Yesilkoy, E. R. Arvelo, Y. Jahani, M. Liu,
  A. Tittl, V. Cevher, Y. Kivshar, and H. Altug, ``Ultrasensitive
  hyperspectral imaging and biodetection enabled by dielectric
  metasurfaces,'' 
  Nature Photonics {\bf 13}, 390-396 (2019). 

\bibitem{foley14} J. M. Foley, S. M. Young,  and J. D. Phillips,
``Symmetry-protected mode coupling near normal incidence for narrow-band
transmission filtering in a dielectric grating,'' \prb\ {\bf 89},
165111 (2014). 

\bibitem{cui16} X. Cui, H. Tian, Y. Du, G. Shi, and Z. Zhou, ``Normal 
  incidence filter using symmetry-protected modes in dielectric 
  subwavelength gratings,'' Sci. Rep. {\bf 6}, 36066 (2016) 

\bibitem{han19} S. Han, L. Cong, Y. K. Srivastava, B. Qiang,
  M. V. Rybin, 
  A. Kumar, R. Jain, W. X. Lim, V. G. Achanta, S. S. Prabhu,
  Q. J. Wang, Y. S. Kivshar, and R. Singh, 
  ``All-dielectric active terahertz photonics driven by bound states
  in the continuum,'' Advanced Materials, {\bf 31}, 1901921 (2019). 

\bibitem{lijun16pra} 
L. Yuan and Y. Y. Lu, 
``Diffraction of plane waves by a periodic array of nonlinear circular
cylinders,''
\pra\  {\bf 94}, 013852 (2016). 


\bibitem{mocella15} V. Mocella and S. Romano, ``Giant field 
  enhancement in photonic lattices,'' \prb\ {\bf 92}, 155117 
  (2015). 

\bibitem{yoon15} J. W. Yoon, S. H. Song, and R. Magnusson, ``Critical 
  field enhancement of asymptotic optical bound states in the 
  continuum,'' Sci. Rep. {\bf 5}, 18301 (2015). 


\bibitem{shipman05} S. P. Shipman and S. Venakides, ``Resonant
  transmission near nonrobust periodic slab modes,'' Phys. Rev. E {\bf 71},
026611 (2005). 

\bibitem{shipman12} S. Shipman and H. Tu, ``Total resonant
  transmission and reflection by periodic structures,'' SIAM
  J. Appl. Math. {\bf 72}, 216-239 (2012).  


\bibitem{fan02} S. Fan and J. D. Joannopoulos, ``Analysis of guided
  resonances in photonic crystal slabs,'' 
\prb\ {\bf 65}, 235112 (2002).


\bibitem{amgad19} A. Abdrabou and Y. Y. Lu,
``Indirect link between resonant and guided modes on uniform and
periodic slabs,''
\pra\ {\bf 99}, 063818 (2019).

\bibitem{marc91} D. Marcuse, Theory of Dielectric Optical Waveguides,
  2nd ed. (Academic Press, Boston, 1991). 


\end{thebibliography}
\end{document}